\documentclass[a4paper,fleqn,colorlinks=true,linkcolor=blue,citecolor=blue,urlcolor=blue]{cas-dc}
\usepackage{amsmath}
\usepackage[numbers,sort&compress]{natbib}
\usepackage{doi}
\usepackage{float}

\graphicspath{{./}}

\usepackage{xcolor}
\usepackage[nameinlink,capitalise,noabbrev]{cleveref}
\crefname{figure}{Fig.}{Figs.}
\Crefname{figure}{Figure}{Figures}
\crefname{equation}{Eq.}{Eqs.}
\Crefname{equation}{Equation}{Equations}
\newcommand{\RefA}[1]{#1}
\newcommand{\RefB}[1]{#1}

\begin{document}
\let\WriteBookmarks\relax
\def\floatpagepagefraction{1}
\def\textpagefraction{.001}

\shorttitle{Spiral motion enables simultaneous homogenization and enhanced transport in granular pipe flow}

\shortauthors{W. T. Ku et~al.}  

\title [mode = title]{Spiral motion enables simultaneous homogenization\newline and enhanced transport in granular pipe flow}  
\author[1]{Wing To Ku}
\ead{wing.ku@fau.de}
\credit{Software, investigation, validation, visualization, writing – first draft}
\affiliation[1]{organization={Institute for Multiscale Simulation, Friedrich-Alexander-Universit\"at Erlangen-N\"urnberg},
    addressline={Cauerstra{\ss}e 3}, 
    city={Erlangen},
    citysep={}, 
    postcode={91058}, 
    country={Germany}
    }

\author[1]{Patric M\"uller}[orcid=0000-0002-0090-4486]
\ead{patric.mueller@fau.de}
\credit{methods, Project administration, funding acquisition, writing - review \& editing}

\author[1]{Thorsten P\"oschel}[orcid=0000-0001-5913-1070]
\cormark[1]
\ead{thorsten.poeschel@fau.de}
\cortext[cor1]{Corresponding author}
\credit{conceptualization, methods, writing - review \& editing, supervision}

\begin{abstract}
Density waves in gravity-driven granular pipe flow reduce transport efficiency and produce large fluctuations in the local solid fraction. Helical wall textures suppress these instabilities but at the expense of reduced throughput. Using discrete element simulations (DEM), \RefA{we find that this homogenization is associated with} spiral particle motion induced by the helical wall texture. Quantifying this motion by the mean angular velocity reveals a strong correlation between spiral motion, flow homogenization, and transport performance. Motivated by this insight, we introduce a helical pipe whose centerline follows a helical path while maintaining a constant circular cross section. This design generates stronger spiral motion than conventional helical wall textures. For suitable design parameters, the helical pipe simultaneously enhances flow homogeneity and increases the mean flow velocity beyond that of a plain straight pipe. These results support spiral motion as a useful diagnostic and design variable for simultaneously improving flow homogeneity and transport efficiency within the investigated parameter range.
\end{abstract}




\maketitle

\section{Introduction}
\label{sec:intro}

Gravity-driven granular pipe flow is prone to the formation of density waves, i.e., alternating dense plugs and dilute regions, which reduce transport efficiency and result in inhomogeneous particle flow. Suppressing these instabilities is an important objective in the design of efficient granular conveying systems. Such density waves have been studied experimentally, numerically and theoretically for several decades \cite{poschel_recurrent_1994, moriyama19984, PhysRevE.59.778, strauss2006plug, riethmuller_langevin_1997, liss_density_2002, ellingsenModelDensityWaves2010, lopez_clogging-jamming_2020, muller_particle_2025, Gao2025}. Granular transport through vertical pipes and narrow channels is encountered in numerous industrial processes, including mining, chemical engineering and bulk solids handling \cite{Beverloo1961, Nedderman1982, hadjigeorgiou2005ore, janda2013, LEPENNEC1995279, PhysRevLett.86.71}.

Various approaches have been proposed to suppress density waves. In this work, we use the term \emph{passive flow control} to describe flow modification achieved solely by modifying the conveying pipe, without external actuation or additional energy input. Passive flow-control strategies are attractive because they improve flow behavior without increasing the complexity or energy demand of the conveying process. In contrast, active flow-control methods, such as mechanical vibration \cite{hunt_2023_yrapr-3pb48} or electric fields \cite{chen2001granular}, require external energy input and may not be applicable in all industrial settings.

Previous studies have therefore explored passive flow-control strategies for clogging reduction and flow homogenization \cite{zuriguel_silo_2011, Verbucheln2015, verbucheln_homogenization_2017, muller_particle_2025}. Among these approaches, helical wall textures have been shown to be particularly effective. \citeauthor{Verbucheln2015} \cite{Verbucheln2015, verbucheln_homogenization_2017} demonstrated that a helical texture on the inner wall of a vertical pipe substantially suppresses density waves. Their results further suggested an apparent trade-off between flow homogeneity and transport efficiency: increasing the strength of the helical wall texture improves homogenization but reduces the mass flow rate, as reflected by a modified Beverloo relation \cite{Mankoc2007, Beverloo1961, Nedderman1982, sperl2007, alonso-marroquin_beverloo_2021}.

Although helical wall textures have been shown to suppress density waves, \RefB{previous studies focused primarily on the resulting axial solid-fraction distribution and transport rate.} \RefA{The particle-scale motion associated with this homogenization has not been identified systematically.} \RefB{The resulting unresolved relationship between pipe shape, particle motion, flow homogeneity, and transport velocity constitutes the research gap addressed in the present work.}

In this work, we \RefA{use three-dimensional discrete element simulations to test whether spiral particle motion is systematically associated with passive flow homogenization. We quantify the spiral component by the mean angular velocity and compare it with the axial solid-fraction fluctuations and mean transport velocity.} \RefB{For this purpose, we compare a plain reference pipe, a helically textured pipe, and a helical pipe under otherwise identical simulation conditions.}

\RefA{The term mechanism is used here in an operational, particle-resolved sense. We do not attempt to derive a closed hydrodynamic description of the granular flow. Such a derivation would require constitutive relations for stress, dissipation, frictional transport, particle rotation, and the effective boundary conditions at the textured or curved wall. A generally accepted closed description containing these ingredients is not available for the present three-dimensional, dissipative, and frictional flow. Instead, we use controlled DEM comparisons to identify relationships between pipe shape, particle motion, and flow homogenization within the investigated parameter range.}

\RefA{Motivated by the relationship observed for the helically textured pipe,} we introduce a helical pipe whose centerline follows a helical path while maintaining a constant circular cross section. Unlike \RefA{the textured pipe, this shape contains no discrete wall asperities. Within the investigated range of helix periods, the helical pipe produces stronger spiral particle motion and can} simultaneously improve flow homogeneity and transport efficiency.

\RefB{The following sections describe the numerical model and flow characterization, present the results for the two passive flow-control concepts, and summarize the principal conclusions.}

\section{Simulation setup and numerical model}
\label{sec:setup}
\subsection{Granular flow through passive flow-control pipes}
\label{sec:pipe}

\begin{figure}[pos=t]
\centering
\includegraphics[width=0.4\columnwidth]{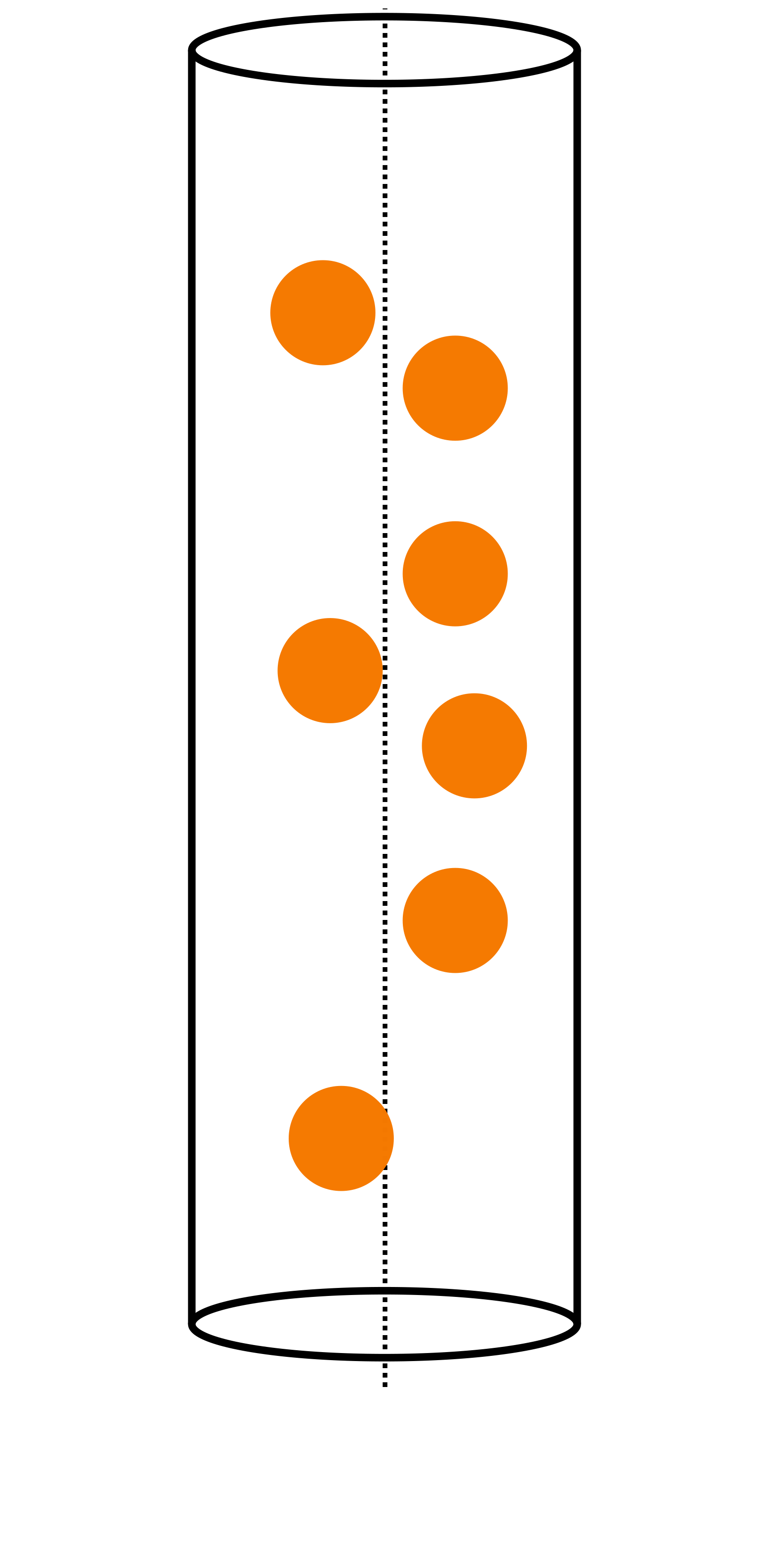}\hspace*{-0.5cm}
\includegraphics[width=0.4\columnwidth]{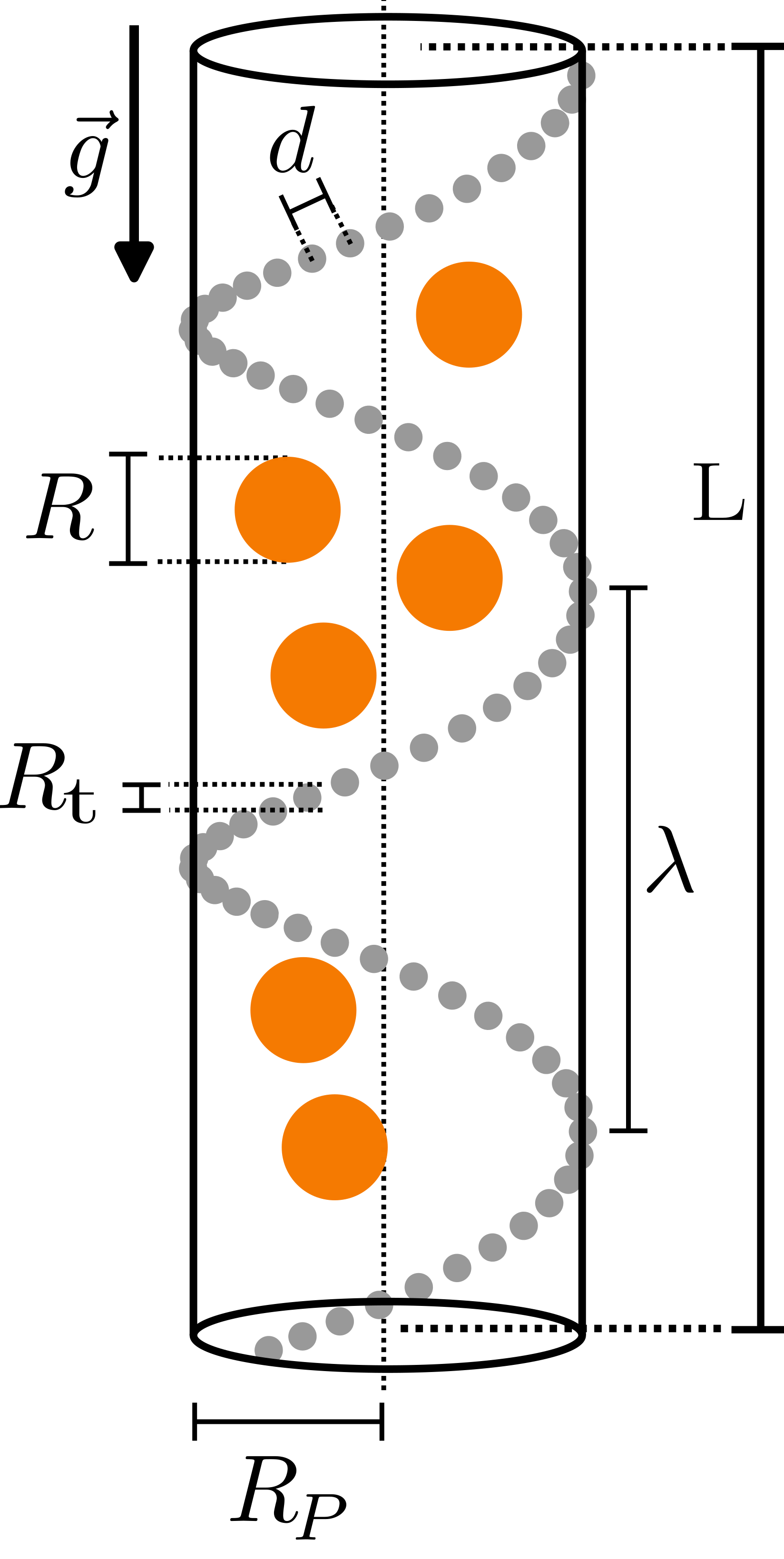}\hspace*{-0.5cm}
\includegraphics[width=0.4\columnwidth]{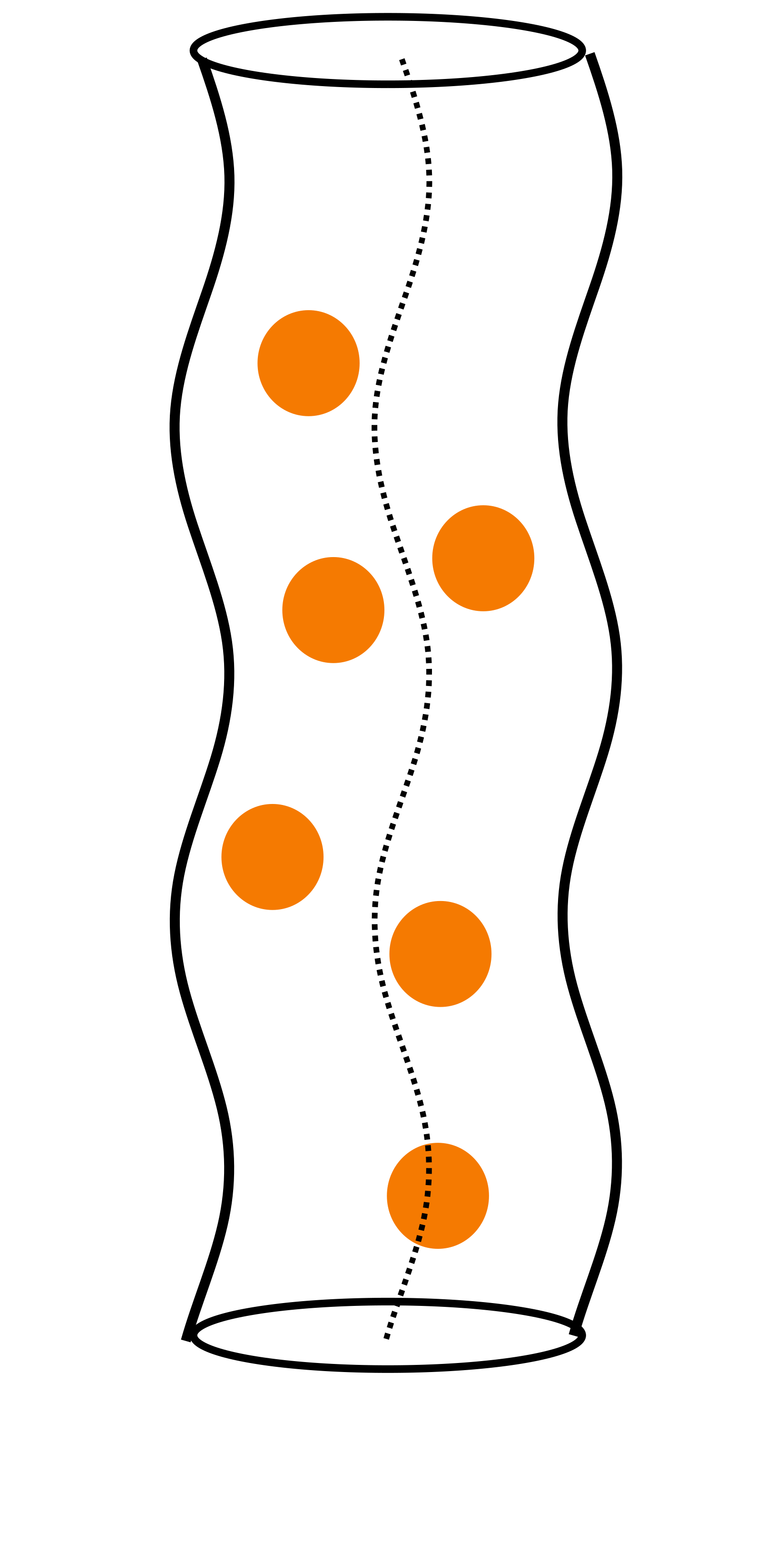}
	\caption{Passive flow-control concepts investigated in this work. From left to right: plain reference pipe, pipe with a helical wall texture, and helical pipe whose centerline follows a helical path while maintaining a constant circular cross section.}
    \label{fig:simSetup}
\end{figure}

We consider gravity-driven flow of spherical particles through a vertical pipe of radius $R_\text{p}$ and length $L$. The simulations consider a plain reference pipe without passive flow-control modifications, a helically textured pipe, and a helical pipe whose centerline follows a helical path while maintaining a constant circular cross section
(\cref{fig:simSetup}). 
The influence of the respective helix period is investigated systematically.

The helical wall texture consists of fixed spherical wall elements of radius $R_\text{t}$ arranged along a helix with period $\lambda$ and spacing $d$, attached to the inner pipe wall. In the helical pipe configuration, the pipe centerline follows a helical path while the pipe maintains a constant circular cross section. Unless stated otherwise, all configurations share identical pipe dimensions and operating conditions.

The flowing particles are \RefA{monodisperse spheres with radius $R$. The use of spherical particles isolates the influence of pipe shape from additional particle-shape effects. In particular, the present model does not include shape-induced interlocking or the enhanced rotational resistance that can occur for angular or elongated particles.}
Initially, $N$ particles are distributed randomly inside the pipe such that the prescribed solid fraction
\begin{equation}
f=\frac{4}{3}\frac{NR^3}{L R_\text{p}^2}
\end{equation}
is obtained. Each particle is assigned a small random radial velocity in the interval $[-v_0,v_0]$. Particle motion is driven solely by gravity acting in the negative $z$ direction, where the $z$-axis is the vertical axis of the simulation domain. Periodic boundary conditions are imposed in the $z$ direction, resulting in statistically stationary flow after an initial transient.
\RefB{All simulations consider a vertical pipe under terrestrial gravity. Pipe inclination and gravity level are not varied, and the conclusions are therefore restricted to this configuration.}

Unless stated otherwise, the simulation parameters listed in \cref{table:param} are used throughout this work.
\begin{table}[pos=t]
\caption{\label{table:param} Reference simulation parameters}
\centering
\begin{tabular}{ lll}
	\hline
	parameter &  & value  \\ \hline	
	particle material density & $\rho$ & \RefA{$900\,\text{kg/m}^3$} \\
    \RefB{gravitational acceleration} & \RefB{$g$} & \RefB{$9.81\,\mathrm{m/s^2}$}\\
	Young's modulus of the particles & $Y$ & 0.1\,GPa\\
	Poisson ratio of the particles & $\nu$ & 0.24\\
    \RefA{dissipative constants} & \\
    \RefA{~~~particle-particle normal} & \RefA{$A_{n,pp}$} &  \RefA{$4.80\times 10^{-6}\,\mathrm{s}$}\\
    \RefA{~~~particle-wall normal} & \RefA{$A_{n,pw}$} & \RefA{$5.51\times 10^{-6}\,\mathrm{s}$}\\
    \RefA{~~~particle-texture normal} & \RefA{$A_{n,pt}$} &\RefA{ $1.43\times 10^{-6}\,\mathrm{s}$}\\
    \RefA{~~~particle-particle tangential} & \RefA{$A_{t,pp}$} & \RefA{$6.31\times 10^{2}\,\mathrm{Pa\,s}$}\\
    \RefA{~~~particle-wall tangential} & \RefA{$A_{t,pw}$} & \RefA{$7.25\times 10^{2}\,\mathrm{Pa\,s}$}\\
    \RefA{~~~particle-texture tangential} & \RefA{$A_{t,pt}$} & \RefA{$1.88\times 10^{2}\,\mathrm{Pa\,s}$}\\
\RefA{particle-particle friction coefficient} & \RefA{$\mu_\mathrm{pp}$} & \RefA{0.5}\\
\RefA{particle-wall friction coefficient} & \RefA{$\mu_\mathrm{pw}$} & \RefA{0.5}\\
	radius of flow particles & $R$ & 0.6\,mm \\ 
	solid fraction of flow particles & $f$ & 0.13\\
    pipe length (periodic bc) & $L$ & $2000R$\\  
	pipe radius & $R_\text{p}$ & $4R$\\ 
	radius of texture particles & $R_\text{t}$ & $0.25R$ \\
	distance of texture particles & $d$ & $2.2R_\text{t}$\\
    limits of initial random velocity &$v_0$ & 0.1\,m/s\\
	\hline	
\end{tabular}
\end{table}

\subsection{DEM simulation}
\label{sec:DEM}

The particle dynamics are simulated using the Discrete Element Method (DEM), as implemented in MercuryDPM \cite{WEINHART2020107129}. The same contact model is employed for particle--particle and particle--wall interactions, combining a viscoelastic Hertzian normal contact force \cite{brilliantovModelCollisionsGranular1996} with a modified Cundall--Strack tangential contact force \cite{cundallDiscreteNumericalModel1979,kloss_models_2012}. 

The normal interaction force between two particles of radii $R_1$ and $R_2$, located at $\vec{r}_1$ and $\vec{r}_2$, acts along the unit vector $\left(\vec{r}_2-\vec{r}_1\right) / \left|\vec{r}_2-\vec{r}_1\right|$ and has the magnitude
\begin{equation}
f_n=\frac{2Y}{3(1-\nu^2)}\sqrt{R_\text{eff}}\left(\xi^{3/2}+\frac{3}{2}A_n \sqrt{\xi}\,\dot{\xi}\right)\,, 
\label{eq:viscel}
\end{equation}
where $\xi\equiv R_1+R_2-\left|\vec{r}_1-\vec{r}_2\right|$ is the mutual deformation and $R_\text{eff}\equiv R_1R_2/(R_1+R_2)$. The Young's modulus, $Y$, and the Poisson ratio, $\nu$, are material properties specified in \cref{table:param}. The dissipative parameter $A_n$ is determined from the universal relation between the velocity-dependent coefficient of restitution and the dimensionless impact velocity \cite{brilliantovModelCollisionsGranular1996, Ramirez1999, Schwager1998, Schwager2008}. \RefA{It differs for particle–particle, particle–wall, and particle–texture collisions.} Numerically inverting this expression \cite{mullerCollisionViscoelasticSpheres2011} and assuming the coefficient of restitution of $0.5$ at an impact velocity of $1\,\text{m/s}$, we obtain the values given in \cref{table:param}.

The tangential interaction force acts in the direction of the projection of the relative particle velocity onto the contact plane and is given by
\begin{equation}
f_t=\min\left(\mu f_n,\frac{4G}{2-\nu}\sqrt{R_\text{eff}\xi}\int_0^\tau v_t dt +A_t\sqrt{R_\text{eff}\xi}v_t\right)\,,
\label{eq:ft}
\end{equation}
where $\mu$ \RefA{denotes either $\mu_\mathrm{pp}$ or $\mu_\mathrm{pw}$,
depending on the interacting pair,} $v_t$ is the relative tangential velocity at the contact point, and
\begin{equation}
    G \equiv \frac{Y}{2(1+\nu)}\qquad\text{and}\qquad A_t=A_n\frac{Y}{1-\nu}
\end{equation}
denote the shear modulus and the tangential dissipative parameter \cite{muller_particle_2025}, respectively. \RefA{Thus, $A_t$ is not an independently selected simulation parameter but follows from $A_n$, $Y$, and $\nu$. As in the case of $A_n$, $A_t$ differs for particle–particle,
particle–wall, and particle–texture collisions. The term containing $A_t$ describes the dissipative contribution to the tangential contact force.} The integral in \cref{eq:ft} is performed over the duration of the contact, $\tau$. 

\RefA{The Young's modulus $Y=0.1\,\mathrm{GPa}$ corresponds to hard rubber or plastics like low-density polyethylene. All pipe configurations are compared using the same Young's modulus and contact model. The comparison therefore isolates the influence of pipe shape under identical numerical stiffness conditions. However, no systematic stiffness sensitivity study was performed, and the present results do not demonstrate that the density-wave dynamics are independent of $Y$.} 

\RefA{The same Coulomb friction coefficient, $\mu_\mathrm{pp}=\mu_\mathrm{pw}=0.5$, is used for particle--particle and particle--wall contacts. Because particle--wall friction directly affects the transfer of angular momentum from the wall to the particle flow, the quantitative results may depend on both friction coefficients. No systematic friction-sensitivity study was performed; the conclusions are therefore restricted to the values reported here.}

The equations of motion are integrated using a time step of $\Delta t = 6\times 10^{-7}\,\text{s}$. 
\RefA{For the reference parameters, the Rayleigh time \cite{Otsubo2017} is estimated as
\begin{equation}
t_\mathrm{R} =
\frac{\pi R\sqrt{\rho/G}} {0.1631\nu+0.8766}\,.
\end{equation}
Using the parameters listed in \cref{table:param}, we obtain $t_\mathrm{R}=9.73\times10^{-6}\,\mathrm{s}$ and therefore $\Delta t/t_\mathrm{R}=0.062$. The integration time step is thus substantially smaller than the Rayleigh time and satisfies the commonly
used stability criterion for DEM simulations.}  \RefA{This estimate verifies the stability of the selected time step but does not constitute a time-step independence study. A systematic time-step refinement study was not performed for the present data set.}

\RefA{All particle positions, velocities, and forces are evaluated in the inertial laboratory frame. The equations of motion contain gravity and the particle-particle and particle-wall contact forces described above. The influence of the pipe shape enters through the particle-wall contact forces. No rotating or centerline-following reference frame is introduced; consequently, no centrifugal or Coriolis forces appear in the equations of motion.}

\subsection{Flow characterization}
\label{sec:flow}

To quantify the effect of passive flow control, we evaluate three quantities: the mean flow velocity, the standard deviation of the local solid fraction, and the mean angular velocity of the particle flow. The first two characterize transport efficiency and flow homogeneity, whereas the latter quantifies the intensity of spiral particle motion.

The mean flow velocity is obtained by averaging the axial particle velocity over all flowing particles, followed by a time average after the flow has reached a statistically stationary state.

Flow homogeneity is quantified by the standard deviation of the local solid fraction. To this end, the pipe is divided into $N_\text{part}=100$ sections of equal length along the pipe axis. The local solid fraction of section $s$ is computed as
\begin{equation}
f_s=\frac{N_\text{part}}{\pi R_\text{p}^2L} \sum_{i=1}^{N} V_i^s\,,
\end{equation}
where $V_i^s$ denotes the volume of particle $i$ contained within section $s$. The standard deviation of $f_s$ is then calculated and averaged over time after the flow has reached a statistically stationary state.
\RefB{Because the simulations employ periodic boundary conditions, they have no inlet or outlet from which a conventional residence-time distribution could be determined. Furthermore, all particles have the same size and material properties, so particle segregation is not defined for the present monodisperse system. The local solid-fraction measure instead quantifies axial flow uniformity throughout the periodic domain.}

To quantify the intensity of spiral particle motion, we evaluate the mean angular velocity about the vertical pipe axis,
\begin{equation}
\omega =
\left|
\frac{
\sum_{p=1}^{N}
\left(\vec{r}_{p,\perp}\times\vec{v}_p\right)
\cdot\vec{e}_z
}{
\sum_{p=1}^{N}|\vec{r}_{p,\perp}|^2
}
\right|\,,
\end{equation}
where $\vec{e}_z\equiv(0,0,1)$ and
$\vec{r}_{p,\perp}
=\vec{r}_p-(\vec{r}_p\cdot\vec{e}_z)\vec{e}_z$
is the radial position of particle $p$ relative to the vertical pipe
axis.

\RefA{The mean angular velocity $\omega$ is a kinematic diagnostic and
is not itself a conserved quantity. The total angular momentum of the
particle system about the vertical axis is
\begin{equation}
L_z =
\sum_{p=1}^{N}
\left[
m_p\left(\vec{r}_p\times\vec{v}_p\right)
+I_p\vec{\Omega}_p
\right]\cdot\vec{e}_z\,,
\end{equation}
where $I_p$ and $\vec{\Omega}_p$ denote the moment of inertia and spin
angular velocity of particle $p$. Its balance is
\begin{equation}
\frac{\mathrm{d}L_z}{\mathrm{d}t}
=
\sum_{c\in\mathrm{wall}}
\left(\vec{r}_c\times\vec{F}_c\right)\cdot\vec{e}_z\,,
\end{equation}
where $\vec{r}_c$ is the position of wall contact $c$ and $\vec{F}_c$
is the corresponding force acting on the particle. Gravity produces
no torque about the vertical axis, and particle--particle contact
contributions cancel in the balance of the complete particle system.
Thus, particle--wall contacts provide the external torque, while the
wall shape determines its axial component. The present simulations
quantify the resulting spiral motion through $\omega$; a
contact-resolved quantitative torque budget was not evaluated.}

\section{Results}
\subsection{Flow homogenization by helical wall textures}
\label{sec:homo}

\begin{figure}[pos=b]
\centering
\includegraphics[width=1.0\columnwidth]{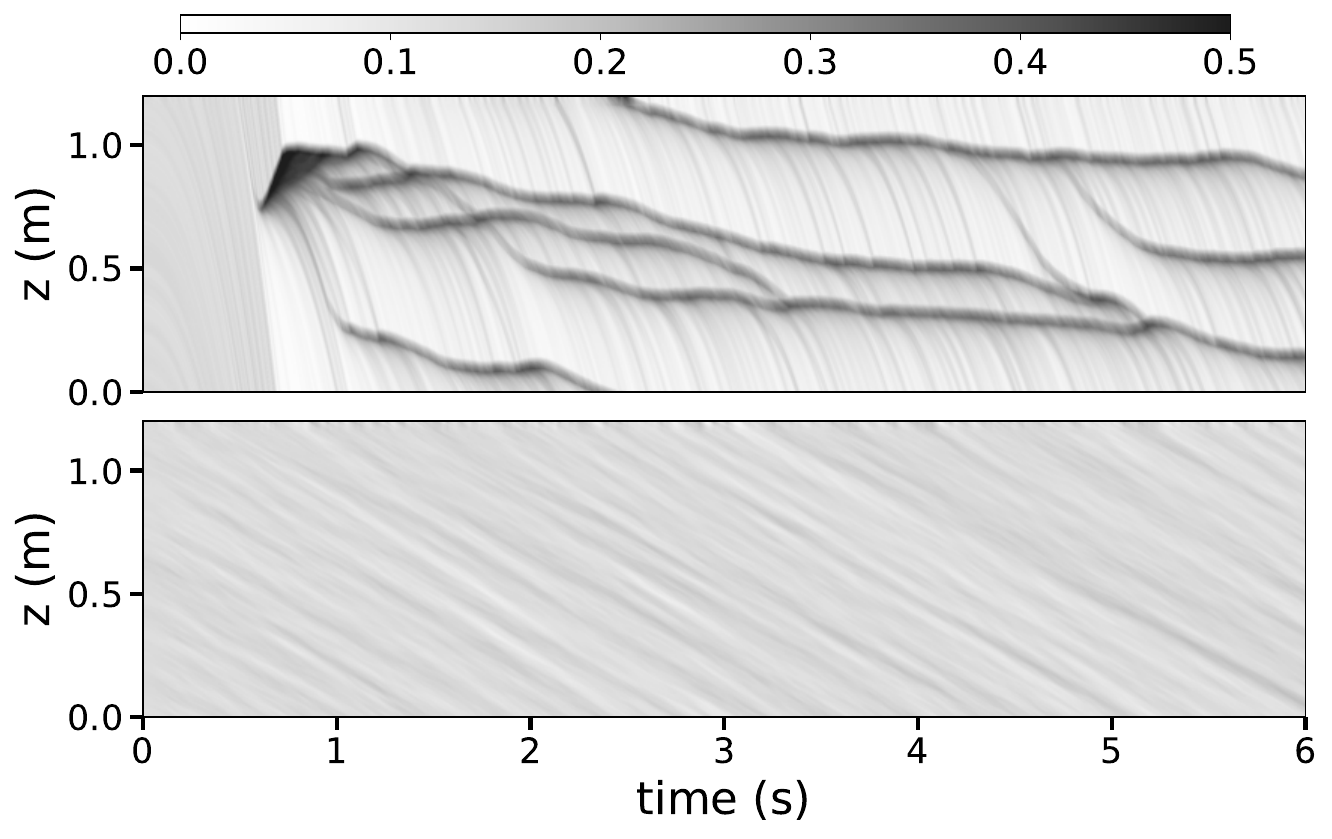}
	\caption{Spatio-temporal plots of the solid fraction along
the pipe. Upper panel: formation of plugs in the plain reference pipe;
lower panel: homogenization due to helical inner wall texture; The gray value indicates the packing fraction.}
    \label{fig:solidfraction_custom_plot}
\end{figure}

\Cref{fig:solidfraction_custom_plot} 
compares the spatio-temporal evolution of the local solid fraction in the plain reference pipe and a pipe with a helical wall texture. In the plain pipe, pronounced density waves develop in the form of alternating dense plugs and dilute regions. These plugs propagate along the pipe and produce large temporal and spatial fluctuations of the local solid fraction. In contrast, the helical wall texture largely suppresses the formation of dense plugs, resulting in a considerably more homogeneous particle flow.

The pronounced difference between the two flow patterns demonstrates that a helical wall texture effectively suppresses density-wave formation in gravity-driven granular pipe flow. While this homogenizing effect has previously been reported \cite{Verbucheln2015, verbucheln_homogenization_2017}, the physical mechanism responsible for it has remained unclear. Since the wall texture imposes a helical path on particles moving close to the wall, a natural hypothesis is that the induced spiral particle motion governs the observed homogenization. In the following section, we examine this hypothesis by systematically relating the intensity of spiral motion to both flow homogeneity and transport performance.

\subsection{\RefA{Spiral motion and flow homogenization}}
\label{sec:spiral-govern}

\begin{figure}[pos=htb]
\centering
\includegraphics[width=0.8\columnwidth]{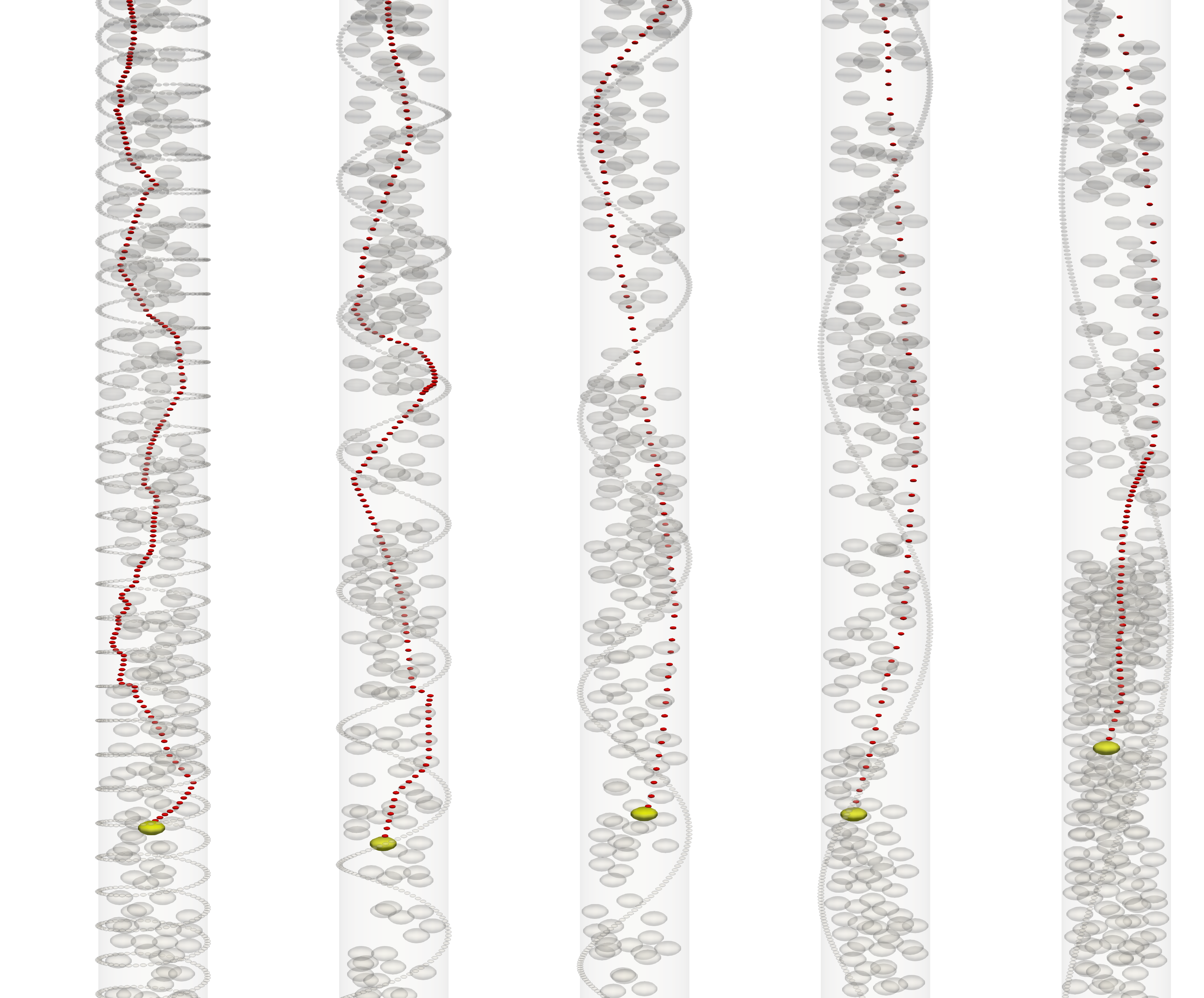}
\caption{Effect of wall texture on the flow pattern. The trajectory of a particle (yellow) is highlighted in red. From left to right: $\lambda = 5R, 20R, 40R, 80R$ and $125R$. As $\lambda$ increases, particles develop more spiral motion, but the spiral motion decays as plugs form, at $\lambda = 125R$. The particles appear ellipsoidal because the scale in the direction of gravity is compressed by a factor of 2 to visualize a longer section of the pipe. 
}
    \label{fig:texture_flowpattern}
\end{figure}

To investigate the role of spiral motion in passive flow control, we systematically vary the helix period $\lambda$ of the wall texture. Figure \ref{fig:texture_flowpattern} shows typical flow patterns in the pipe with helix wall texture different periods. At  $\lambda = 5R$,  particles flow with very limited spiral motion, and the flow hardly follows the periodicity of the helical wall texture. As $\lambda$ increases to $\lambda=80R$, the flow develops more spiral motion. Although the flow still spirals at a longer period than that of the corresponding wall texture, the particle trajectory appears smoother. As  $\lambda$ further increases to $\lambda=125R$, plugs form, hindering the spiral motion.

\begin{figure}[pos=htb]
\centering
\includegraphics[width=0.7\columnwidth]{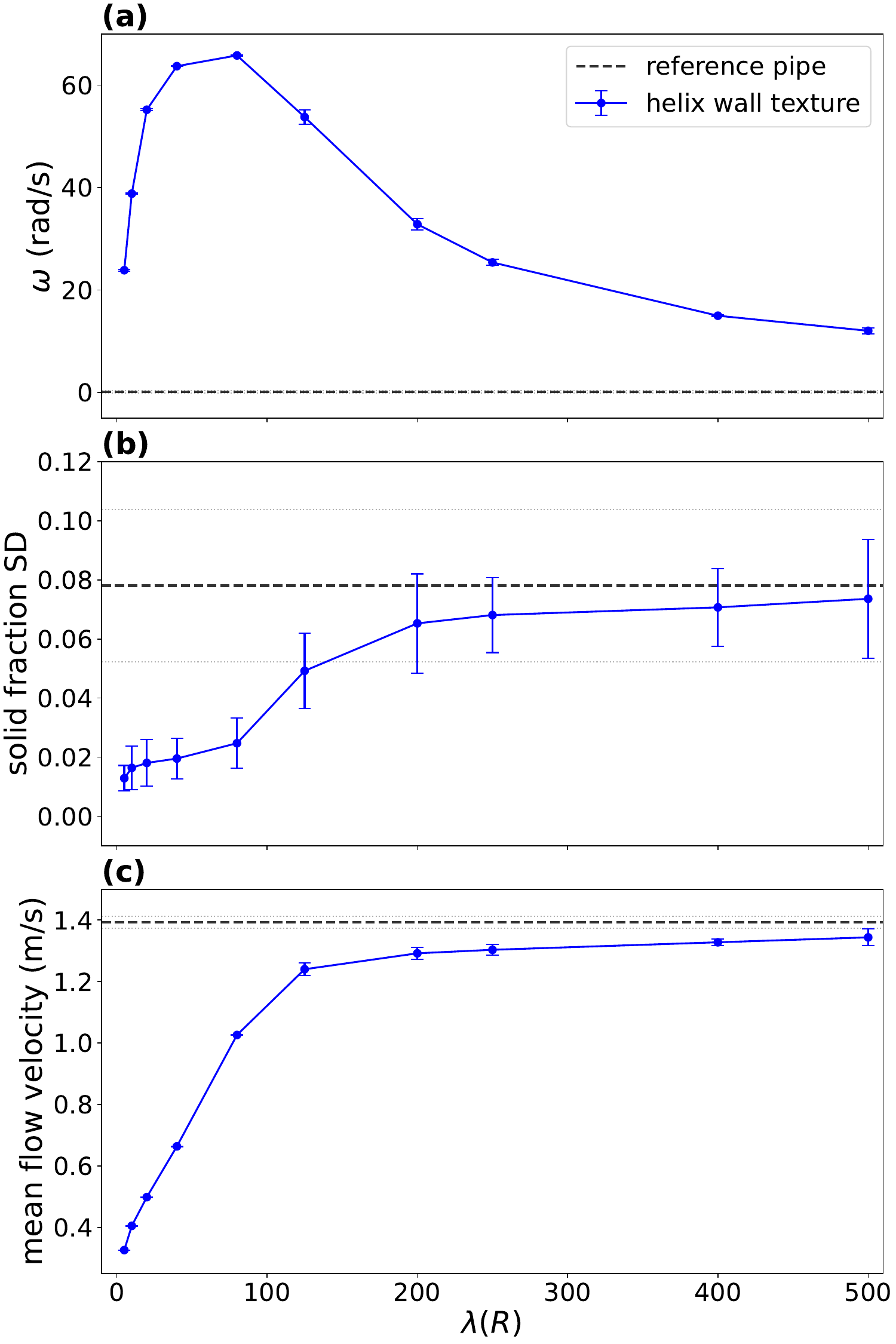}
\caption{Effect of the helix period on spiral motion, flow homogeneity, and transport performance. (a) Mean angular velocity. (b) Standard deviation of local solid fraction. (c) Mean flow velocity. Error bars denote one standard deviation obtained from 5 independent simulations. The dashed lines indicate the corresponding values and error bars for the reference pipe.}
    \label{fig:walltexture}
\end{figure}

For each helix period, the mean angular velocity, the standard deviation of the local solid fraction, and the mean flow velocity are evaluated as described in \cref{sec:flow}. \Cref{fig:walltexture} summarizes the dependence of these quantities on the helix period. Each data point shown represents the mean and standard deviation obtained from 5 independent simulations with different random initial particle positions and velocities.

\Cref{fig:walltexture}a shows that the mean angular velocity strongly depends on $\lambda$. As the helix period increases, the angular velocity first rises, reaches a maximum for $\lambda \approx 80R$, and decreases for larger helix periods. Thus, the helix period directly controls the intensity of the induced spiral motion.

The standard deviation of the local solid fraction exhibits the opposite trend (\Cref{fig:walltexture}b). It remains small for helix periods up to $\lambda \approx 80R$, indicating a homogeneous particle flow, but increases rapidly for larger helix periods as density waves reappear. The strong inverse correlation between the standard deviation of the local solid fraction and the mean angular velocity \RefA{shows that stronger spiral motion is consistently associated with the} suppression of density waves over the investigated range of helix periods.

The mean flow velocity (\Cref{fig:walltexture}c) initially increases with increasing helix period, reaches a maximum near $\lambda \approx 125R$, and subsequently approaches the value of the reference pipe. Although strong spiral motion promotes flow homogenization, it does not lead to the highest transport velocity. Instead, the maximum flow velocity remains below that of the reference pipe.

These results \RefA{support} spiral particle motion as a particle-scale feature \RefA{associated with} flow homogenization in helically textured pipes. \RefA{Stronger spiral motion is accompanied by reduced} density-wave formation, whereas the wall texture used to generate this motion also increases particle-wall interactions and the associated energy dissipation. The transport performance therefore results from a competition between these two effects. Consequently, an optimal helix period exists at which the flow is effectively homogenized while the additional flow resistance remains small. This finding motivates the search for alternative passive flow-control concepts that generate spiral motion more efficiently. A comparison with randomly distributed wall elements is included in the Appendix, \Cref{app:random}. 

\subsection{Helical pipe}
\label{sec:helical-pipe}

The previous section shows that spiral particle motion effectively suppresses density waves, but that the helical wall texture required to generate this motion also introduces additional flow resistance. This suggests that improved passive flow control may be achieved by generating spiral motion without modifying the pipe wall. We therefore investigate a helical pipe whose centerline follows a helical path while maintaining a constant circular cross section.
\begin{figure}[pos=ht]
\centering
\includegraphics[width=0.8\linewidth]{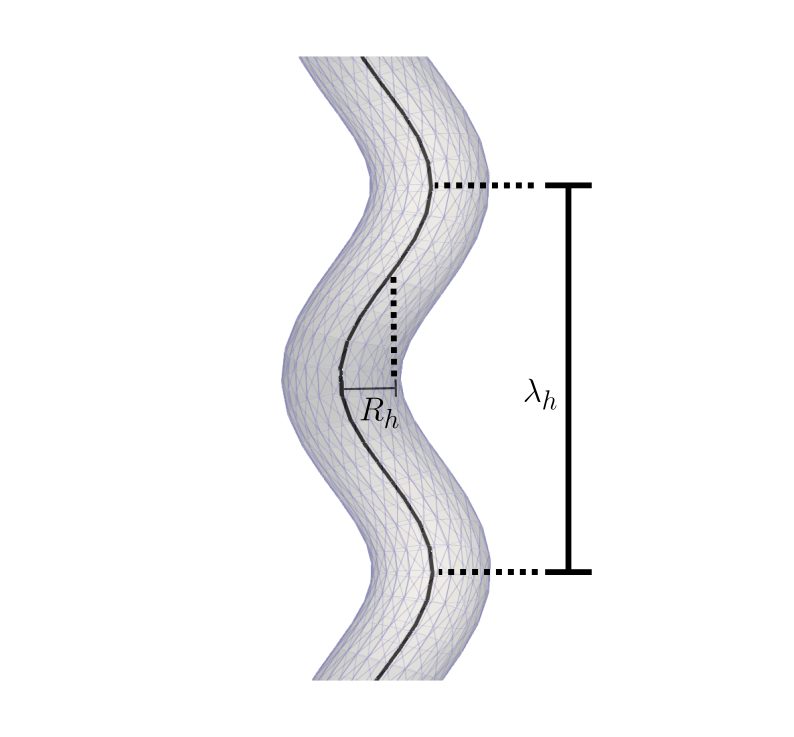}
\caption{Three-dimensional CAD model of a helical pipe whose centerline follows a helical path while maintaining a constant circular cross section. The geometric parameters defining the helical pipe are: helix radius $R_\text{h}$ and helix period $\lambda_\text{h}$.}
 \label{fig:helixspine_schem}
\end{figure}
\Cref{fig:helixspine_schem} illustrates the pipe design. The shape of the pipe is characterized by the helix radius $R_\text{h}$ and the
helix period $\lambda_\text{h}$.
\RefB{By construction, the helix radius equals $\frac{3}{4}$ of the pipe radius, $R_\text{h}=\frac{3}{4} R_\text{p}$, and is therefore not an independent parameter of the present pipe design. The geometric parameter varied in this study is the helix period $\lambda_\text{h}$; the pipe diameter and particle properties are kept constant.}

The pipe shape is generated using a computer-aided design (CAD) model and imported into MercuryDPM as triangulated walls. Unless stated otherwise, the same material parameters, contact model, and operating conditions as described in \Cref{sec:setup} are used. To ensure periodicity, the helix period is chosen such that the pipe height is an integer multiple of $\lambda_\text{h}$.

\RefB{DEM does not employ a computational volume mesh. However, the
curved wall of the helical pipe is represented by a triangulated
surface. A systematic triangulation-refinement study was not performed,
and numerical independence with respect to the wall discretization is
therefore not claimed.}

\begin{figure}[pos=htb]
\centering
\includegraphics[width=0.8\columnwidth]{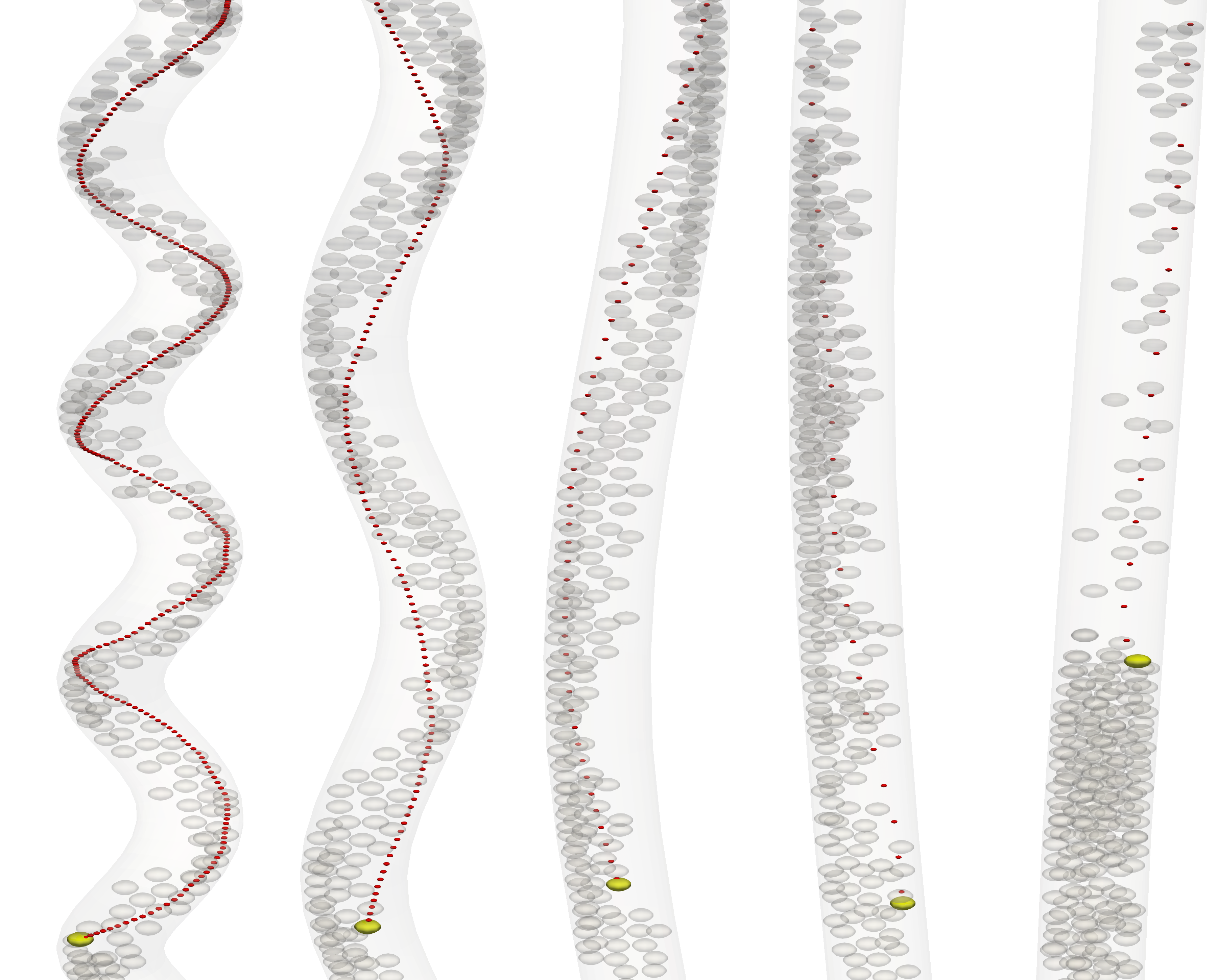}
\caption{Effect of the helical wall on the flow pattern. The trajectory of a particle (yellow) is highlighted in red. From left to right: $\lambda_\text{h} = 40R, 80R, 200R, 400R$ and $500R$.  The particle motion approximately follows the helical period of the pipe, but the spiral motion decays as plugs form, e.g. at $\lambda_\text{h} = 500R$. The particles appear ellipsoidal because the scale in the direction of gravity is compressed by a factor of 2 to visualize a longer section of the pipe. } 
    \label{fig:spine_flowpattern}
\end{figure}

\Cref{fig:spine_flowpattern} shows typical flow patterns in the helical pipe at different periods. In contrast to the helically \RefA{textured pipe, the helical pipe contains no discrete wall asperities. The observed spiral motion results from particle interactions with the smooth curved wall represented in the DEM model.} The resulting spiral motion is visibly stronger and extends across the entire pipe cross section.

\subsection{Comparison of passive flow-control concepts}
\label{sec:comparison}

\begin{figure}
\centering
\includegraphics[width=0.8\columnwidth]{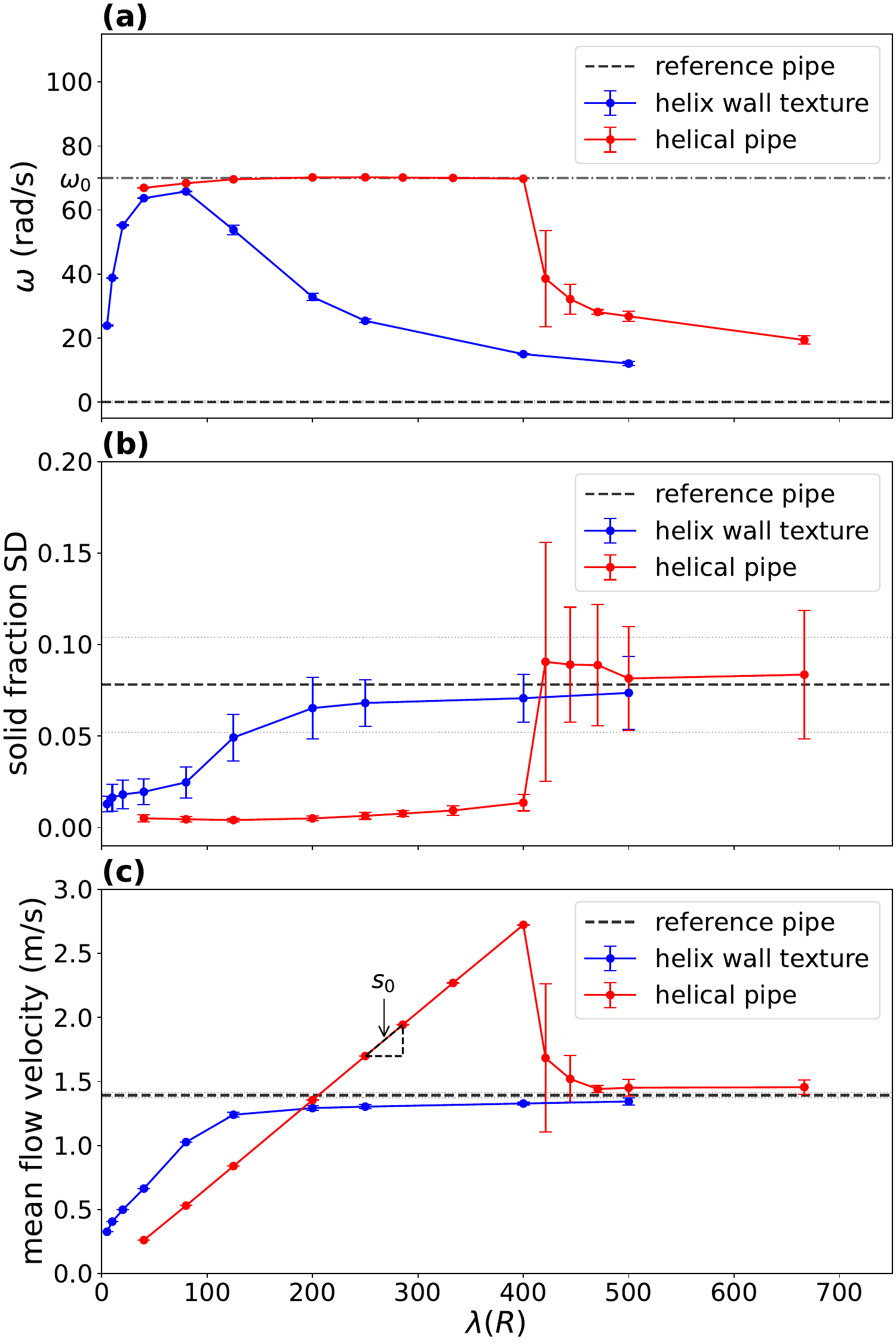}
\caption{Comparison of the helical pipe and the helical wall texture as passive flow-control concepts. (a) Mean angular velocity. (b) Standard deviation of the local solid fraction. (c) Mean flow velocity. The horizontal axis denotes the helix period of the respective passive flow-control concept.  Error bars denote one standard deviation obtained from 5 independent simulations.} 
    \label{fig:spinehelix}
\end{figure}

Motivated by the previous results, we compare the performance of the helical pipe with that of the helical wall texture. \Cref{fig:spinehelix} summarizes the mean angular velocity, the standard deviation of the local solid fraction, and the mean flow velocity for both passive flow-control concepts as functions of the helix period.
\Cref{fig:spinehelix}a shows that the helical pipe generates a greater angular velocity than the helical wall texture, particularly for helix periods in the range $\lambda_\text{h}\in (125R, 400R)$. While the helical wall texture fails to maintain spiral particle motion as the helix period increases, modifying the pipe centerline continues to induce strong spiral motion.

The stronger spiral motion is accompanied by a pronounced reduction in the standard deviation of the local solid fraction (\Cref{fig:spinehelix}b). Throughout the range of helix periods for which the angular velocity is increased, the helical pipe consistently produces a more homogeneous particle flow than the helical wall texture. These results further support the conclusion that spiral particle motion \RefA{is closely related to} flow homogenization.

The corresponding mean flow velocity is shown in \Cref{fig:spinehelix}c. \RefB{For the simulated helix periods $250R\leq\lambda_\mathrm{h}\leq400R$, the helical pipe simultaneously produces a lower mean solid-fraction standard deviation and a higher mean flow velocity than the straight reference pipe. This interval is the demonstrated operating range for simultaneous improvement within the investigated parameter set; it should not be interpreted as a universal design range.} The maximum flow velocity is obtained for $\lambda_\mathrm{h}\approx400R$.

\RefB{Because the gravitational acceleration and the prescribed particle--wall friction coefficient are identical in all simulations, the velocity increase cannot result from changes in these parameters. As explained in \cref{sec:DEM}, the calculations are performed in the inertial laboratory frame and contain no centrifugal force. The increase is associated with modified particle--wall interactions and reduced plug formation, although a quantitative energy and collision budget was not evaluated.}

For the helical pipe, the nearly constant angular velocity and the linear dependence of the mean flow velocity on $\lambda_\text{h}$ (for $50R \le \lambda_\text{h} \le 400R$) indicate that particles rotate in sync with the helix period. Extracting the constant $\omega_0 \approx 70.0\,\mathrm{rad/s}$ in the angular velocity and the slope  $s_0 \approx 11.4/s $ in the mean flow velocity from \Cref{fig:spinehelix}, we compute the spatial frequency $k$—the number of particle rotations per unit distance traveled:
\begin{equation}
k = \frac{\omega}{2\pi v} \approx  \frac{\omega_0}{2\pi s_0 \lambda_\text{h}} \approx \frac{1}{\lambda_\text{h}}.
\end{equation}
Thus, the particles complete approximately one rotation over each axial distance $\lambda_\text{h}$.

As $\lambda_\text{h}$ increases further above $400R$, the formation of plugs disrupts the flow pattern, leading to a reduction in angular velocity. At the same time, the mean flow velocity decreases, while the standard deviation of the solid fraction increases toward the value observed in the reference pipe.

The comparison demonstrates that the previously observed trade-off between flow homogenization and transport efficiency is not an inherent consequence of spiral particle motion. \RefB{The helically textured pipe is straight and generates spiral motion through discrete wall elements, whereas the helical pipe has a curved centerline and no discrete wall asperities. Because both configurations use the same contact parameters and particle--wall friction coefficient, the comparison shows that pipe curvature can generate spiral motion and simultaneous performance improvement without helical wall roughness.} The present data do not, however, provide a quantitative decomposition of all curvature- and collision-related contributions.

\RefB{Particle--wall collision frequency, impact intensity, and wall wear were not evaluated. Stronger transverse particle motion may increase wall loading, and this practical trade-off must be examined before industrial application. Computational-cost and scaling data were likewise not retained. Although all configurations use the same number of particles and contact model, the helical pipe employs a triangulated CAD wall, whereas the straight and helically textured pipes use simpler wall representations. Quantitative computational benchmarking, wear analysis, and experimental validation therefore remain necessary.}

\section{Conclusion}
\label{sec:conclusion}

In this work, we investigated the mechanism by which passive flow-control concepts suppress density waves in gravity-driven granular pipe flow. Using discrete element simulations, we \RefA{found a consistent relationship between spiral particle motion and the suppression of density waves.} 

The analysis of helically textured pipes showed that increasing spiral motion improves flow homogeneity but does not necessarily increase transport performance. Although an appropriately chosen helix period effectively suppresses density waves, the wall texture required to generate spiral motion also increases particle--wall interactions and the associated energy dissipation. Consequently, the transport performance results from a competition between enhanced flow homogenization and the additional resistance introduced by the wall texture.

\RefA{Motivated by this observed relationship,} we introduced a helical pipe whose centerline follows a helical path while maintaining a constant circular cross section. In contrast to the helically \RefA{textured pipe, the helical pipe generates spiral motion through particle interactions with a smooth curved wall rather than with discrete wall asperities.} As a result, substantially stronger spiral motion is achieved over a wide range of helix periods. This leads to both a lower standard deviation of the local solid fraction and a higher mean flow velocity. In particular, the maximum transport velocity exceeds that of both the helical wall texture and the reference pipe.

These findings demonstrate that the previously observed trade-off between flow homogeneity and transport efficiency is not an intrinsic consequence of spiral particle motion. Instead, it arises from the specific passive flow-control concept used to generate this motion. The present results \RefA{suggest spiral particle motion as a useful diagnostic and possible design variable} for passive flow control in granular pipe flow. \RefA{Future studies should determine whether passive flow-control concepts can be improved by increasing spiral particle motion while limiting} the accompanying flow resistance.

\appendix
\section*{Appendix}
\section{Comparison with random wall textures}
\label{app:random}

To assess whether the observed flow homogenization is a consequence of the helical arrangement rather than wall roughness alone, we compare the helical wall texture with randomly distributed wall elements. Following \citet{Verbucheln2015}, the random textures are generated using the same number of wall elements as the corresponding helical textures.

\begin{figure}[pos=htb]
\centering
\includegraphics[width=0.8\columnwidth]{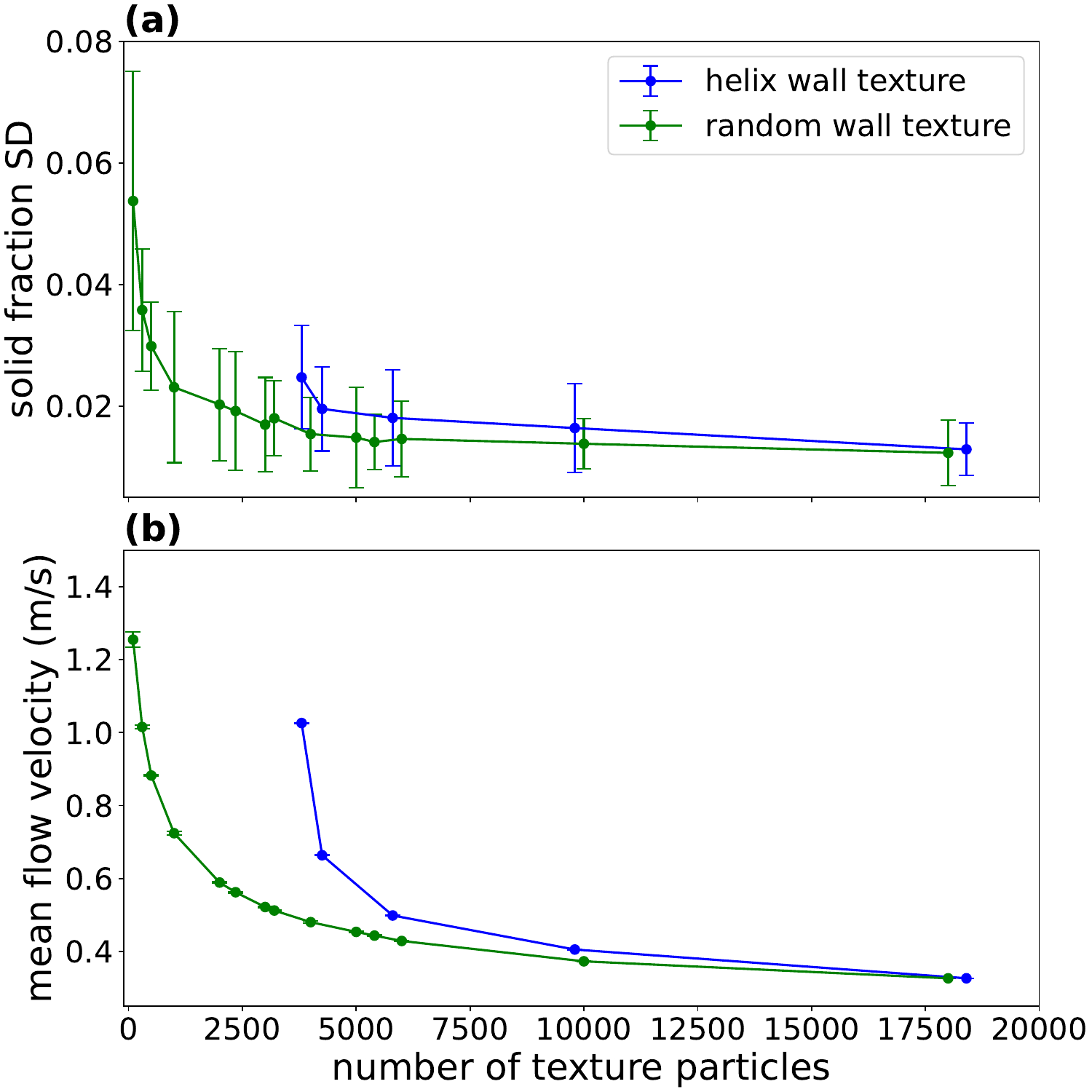}
\caption{Effect of the number of randomly distributed wall
elements on flow homogeneity and transport performance. Top:
Standard deviation of the local solid fraction. Bottom: Mean flow velocity.}
    \label{fig:randomtexture}
\end{figure}

\Cref{fig:randomtexture} 
shows the standard deviation of the local solid fraction and the mean flow velocity as functions of the number of wall elements. When the number of wall elements is greater than $3500$, although the random wall texture maintains a slightly smaller flow homogeneity, it results in a much lower mean flow velocity than the helix wall texture. Reducing the number of randomly distributed wall elements below $3500$ increases the mean flow velocity, but simultaneously deteriorates flow homogeneity. Consequently, random wall textures cannot achieve the combination of efficient transport and effective flow homogenization obtained with helical wall textures.

\section*{Funding}
This work was funded by the Deutsche Forschungsgemeinschaft (DFG, German Research Foundation) -- project number 398618334. The funder had no involvement in the study design, analysis and interpretation of the results, preparation of the manuscript, or decision to submit it for publication.

\section*{Declaration of competing interest}
The authors declare no financial or personal relationships that could have influenced the research presented in this article.

\section*{Data availability}
The simulation data underlying the results presented in this article are available from the corresponding author upon request. 

\printcredits

\section*{Declaration of generative AI and AI-assisted technologies in the manuscript preparation process}
During the preparation of this work, the authors used ChatGPT for English language improvement. The authors reviewed and edited the output as needed and take full responsibility for the content of the published article. 

\bibliographystyle{unsrtnat}

\bibliography{granular-pipe-TP}

\end{document}